# Ultrafast Laser-Induced Melting of Long-Range Magnetic Order in Multiferroic TbMnO$_3$

J. A. Johnson[1,*,†], T. Kubacka[2], M. C. Hoffmann[3], C. Vicario[4], S. de Jong[3], P. Beaud[1], S. Grübel[1], S.-W. Huang[1], L. Huber[2], Y. W. Windsor[1], E. M. Bothschafter[1], L. Rettig[1], M. Ramakrishnan[1], A. Alberca[1], L. Patthey[4], Y.-D. Chuang[5], J. J. Turner[3], G. L. Dakovski[3], W.-S. Lee[3], M. P. Minitti[3], W. Schlotter[3], R. G. Moore[6], C. P. Hauri[4,7], S. M. Koohpayeh[8], V. Scagnoli[1,9], G. Ingold[1], S. L. Johnson[2], U. Staub[1].

[1]Swiss Light Source, Paul Scherrer Institut, 5232 Villigen PSI, Switzerland.

[2]Institute for Quantum Electronics, Eidgenössische Technische Hochschule (ETH) Zürich, 8093 Zürich, Switzerland.

[3]Linac Coherent Light Source (LCLS), SLAC National Accelerator Laboratory, Menlo Park, CA 94025, USA.

[4]SwissFEL, Paul Scherre Institut, 5232 Villigen PSI, Switzerland.

[5]Advanced Light Source, Lawrence Berkeley National Laboratory, Berkeley, CA 94720, USA.

[6]Stanford Institute for Materials and Energy Sciences (SIMES), SLAC National Accelerator Laboratory, Menlo Park, CA 94025, USA.

[7]Ecole Polytechnique Federale de Lausanne, 1015 Lausanne, Switzerland.

[8]Institute for Quantum Matter (IQM), Department of Physics and Astronomy, Johns Hopkins University, Baltimore, MD 21218, USA.

[9]Laboratory for Mesoscopic Systems, Department of Materials, Eidgenössische Technische Hochschule (ETH) Zürich, 8093 Zürich, Switzerland.

We performed ultrafast time-resolved near-infrared pump, resonant soft X-ray diffraction probe measurements to investigate the coupling between the photoexcited electronic system and the spin cycloid magnetic order in multiferroic TbMnO$_3$ at low temperatures. We observe melting of the long range antiferromagnetic order at low excitation fluences with a decay time constant of 22.3 ± 1.1 ps, which is much slower than the ~1 ps melting times previously observed in other systems. To explain the data we propose a simple model of the melting process where the pump laser pulse directly excites the electronic system, which then leads to an increase in the effective temperature of the spin system via a slower relaxation mechanism. Despite this apparent increase in the effective spin temperature, we do not observe changes in the wavevector $q$ of the antiferromagnetic spin order that would typically correlate with an increase in temperature under equilibrium conditions. We suggest that this behavior results from the extremely low magnon group velocity that hinders a change in the spin-spiral wavevector on these time scales.

# I. Introduction

Correlated materials, and more specifically multiferroics with strong coupling between electronic, structural, magnetic, orbital, charge, and other degrees of freedom, have widely recognized potential for future generations of solid-state computational devices [1]. In addition to possible applications, understanding the strong competition between different degrees of freedom in correlated electron systems poses significant experimental and theoretical challenges. One experimental approach to studying such couplings is ultrafast pump-probe measurements, where a femtosecond-duration light pulse perturbs the material and a variety of time-resolved probes of the system can observe the induced changes and subsequent relaxation dynamics. These measurements can potentially directly measure the strength of coupling between different internal degrees of freedom, and can additionally reveal relevant time scales and control mechanisms for possible applications.

The perovskite manganites, derivatives of $LaMnO_3$, are prominent examples of strongly correlated electron materials that posses rich phase diagrams and exhibit a number of interesting effects such as colossal magnetoresistance. One of these derivatives, $TbMnO_3$, exhibits multiferroicity below 27 K [2], and serves as a prototypical example of chiral spin order. The nature of the long range spin-cycloid ordering under equilibrium conditions has been studied and clarified using both neutron scattering [3-6] and resonant x-ray diffraction [7-9]. In the latter technique, soft X-rays tuned to the Mn $L_2$ edge at 652 eV show a strong resonant enhancement giving direct information about the long-range spin order. At room temperature, $TbMnO_3$ is paramagnetic and no spin-order diffraction peaks are observed. As the temperature is lowered below $T_{N1}$ = ~42 K, the spin system develops a spin density wave state with the material remaining in a paraelectric state. In a resonant X-ray scattering experiment, this results in a well-defined, incommensurate ($0q0$) diffraction peak, where $q \approx$ 0.3 varies slightly with temperature [7]. Below $T_{N2}$ = ~27 K, the spins in the $bc$-plane order into a spin cycloid creating a spontaneous ferroelectric polarization directed along the $c$ axis. At temperatures below 27 K the magnetic-ordering wavevector locks in at $q$ = 0.287 [3,7].

Although these scattering measurements have given valuable insight into the equilibrium magnetic structure, in order to understand the magnetic structure under strongly non-equilibrium conditions we require time-resolved techniques. Time-resolved resonant X-ray diffraction has been used to probe ultrafast changes in magnetic order in several oxides [10-13]. For example, we have recently shown that THz excitation can resonantly drive a magnon, which leads to a tilt of the spin cycloid structure in $TbMnO_3$ due to magnetoelectric coupling [13]. In other materials, such as CuO or $La_{0.5}Sr_{1.5}MnO_4$, it was observed that pumping with 800 nm light leads to an ultrafast suppression of magnetic order in less than 1 ps and that the magnetic phase transformation is delayed by approximately ¼ of the period of the low lying spin gap excitation [10,12]. Such ultrafast changes in magnetization (less than 1 ps) have also been observed in metallic ferromagnets via all optical pump/probe schemes and optical pump/XMCD (X-ray magnetic circular dichroism) probe experiments [14-16].

To explore the coupling between the photoexcited electronic system and the spin cycloid magnetic order in $TbMnO_3$ at low temperatures, we performed time-resolved measurements where we pump the system with ultrafast, near-infrared light pulses and probe the magnetic structure with resonant soft X-ray diffraction.

# II. Experimental Results

Optical pump, soft X-ray resonant diffraction probe experiments were performed at the Soft X-Ray (SXR) beamline of the Linac Coherent Light Source X-ray Free Electron Laser (LCLS FEL) [17], using the Resonant Soft X-ray Scattering endstation [18]. The (010) cut $TbMnO_3$ single crystal was oriented such that the *a* axis was in the horizontal scattering plane. The lattice parameters of TbMnO$_3$ in the Pbnm setting are a = 5.3 Å, b = 5.86 Å, and c = 7.49 Å. The x-ray photon energy was tuned to the Mn L$_2$ absorption edge, which gives enhanced sensitivity to the first-order (0*q*0) spin-cycloid reflection [7]. The static magnetic ordering has been characterized in detail using the soft x-ray resonant diffraction station RESOXS [19] at the SIM beamline [20] of the Swiss Light Source.

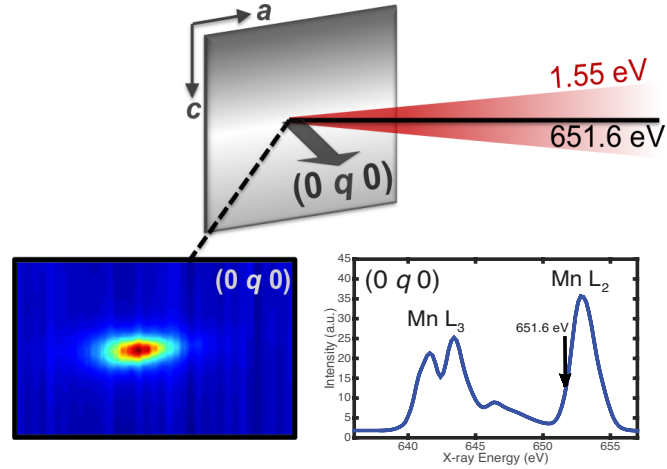

FIG 1. (Color Online) Schematic of experiment. 1.55 eV light pulses are incident on the sample collinear with soft x-ray pulses tuned to the Mn L$_2$ edge (651.6 eV). The (010) cut single crystal TbMnO$_3$ sample azimuth is oriented so that the *a* axis in the scattering plane.

The FEL was operated at a repetition rate of 60 Hz. Kirkpatrick-Baez optics were used to focus the beam to a spot size with a diameter less than 300 μm. Using the fast-CCD camera (fCCD) [18], we measured the (0*q*0) diffraction peak in the in-plane scattering geometry as a function of relative delay between collinear propagating x-ray pulses (π-polarized; 100 fs FWHM) and 800 nm (1.55 eV) pump pulses (~120 fs pulse duration; 118 μm 1/e$^2$ radius; p-polarized).

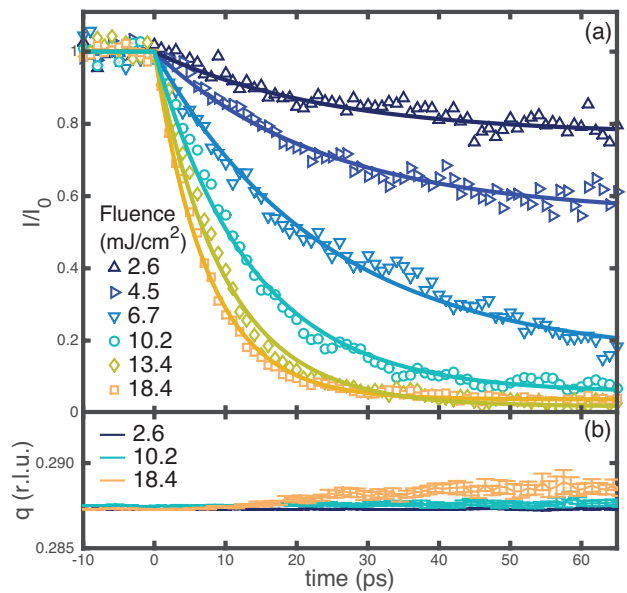

FIG 2. (Color Online) (a) Normalized (0*q*0) diffraction peak intensity as a function of pump probe delay for laser fluences ranging from 2.6 to 18.4 mJ/cm². Standard error bars are roughly the size of the symbols. The dashed lines show fits to a simple exponential form (Eq. 1). (b) Lines with standard errorbars show the measured scattering wavevector as a function of time, for the fluences of 2.6, 10.2, and 18.4 mJ/cm².

Figure 2(a) shows how the diffraction peak intensity varies with time for absorbed pump fluences ranging from 2.6 to 18.4 mJ/cm$^2$, where we have accounted for the reflectivity using dielectric function data from Bastjan *et al.* [21]. As no fast signal components were observed, here we show data binned into one picosecond time-steps. Standard error bars for the peak intensities are roughly the size of the symbols. After 800 nm excitation, the diffraction peak intensity drops. Decay amplitudes and time constants were extracted by fitting the data for $t > 0$ to an exponential function of the form

$$I(t)/I_0 = A[\exp(-t/\tau)-1]+1 . \tag{1}$$

The fit results are shown as dashed lines in Figure 2(a). Decay amplitudes (*A*) and time constants ($\tau$) are shown in Figure 3. For absorbed pump fluences greater than ~6 mJ/cm$^2$, the peak intensity nearly vanishes at long times (see Figure 3(a)), indicating melting of the long-range spin-cycloid order. Residual diffraction peak intensity for even the highest fluence is likely a result of a small misalignment between pump and probe beams and relatively similar spot sizes. At low fluences, the decay time appears to be nearly constant, but then decreases as the decay amplitude saturates with complete melting of the long-range magnetic order; extracted time constants range from 22.4 ps at the lowest fluence to 7.5 ps at the highest fluence, much slower than time constants of <1 ps recorded for melting of magnetic order in other materials [10,12,16]. The decay time is approximately constant for the three lowest fluences, with an average value of $\tau$ = 22.3 ± 1.1 ps. We recorded the diffraction peak position on the fCCD and performed additional theta-2theta scans at various time delays. These measurements all show that the scattering wavevector is nearly constant with time. For the highest fluence, a small shift $\Delta q$ of +0.002 from the nominal value of 0.287 is seen at later times; in Figure 2(b) peak shift data from tracking the peak position on the fCCD are shown by the lines with standard error bars for three fluences (2.6, 10.2, 18.4 mJ/cm$^2$).

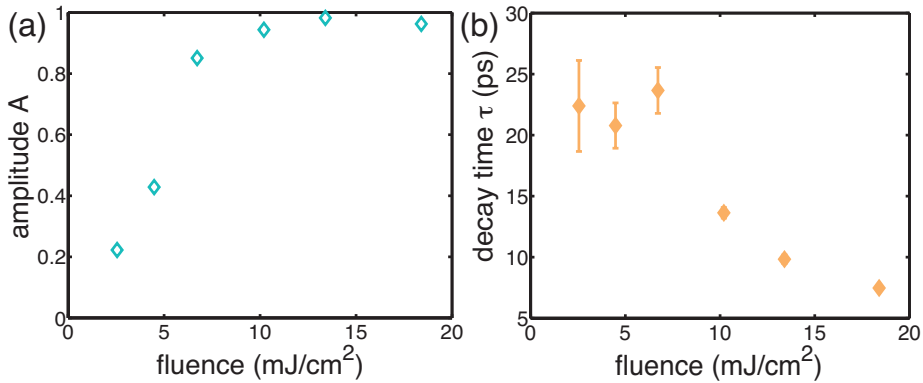

FIG 3. (Color Online) (a) The amplitude of the decay from fits of the data using Eq. (1). An amplitude of 1 indicates complete melting of long-range magnetic order (b) Decay times obtained from fits to Eq. (1) with standard error bars.

### III. Discussion

To better understand the melting of magnetic order we observe in TbMnO$_3$, we first consider the immediate mechanisms of optical absorption. In general the manganites show two prominent, broad features in the imaginary part of the dielectric function in the ultraviolet to near infrared spectral range [21,22]. At higher energies there is a strong increase typically attributed to charge transfer transitions from O 2*p* → Mn 3*d* levels, creating locally an Mn$^{2+}$ excited state. These transitions are centered in the 4-5 eV range with absorption tails extending down to 1 eV. In addition to the charge transfer transitions, there is a broad absorption peak centered close to 2 eV, particularly for light polarized along the ferromagnetically-spin-ordered *a* axis, attributed mainly to intersite Mn$^{3+}$ *d-d* transitions.

This transition results in the transfer of an electron from one Mn site to another, resulting in the creation of $Mn^{4+}$ and $Mn^{2+}$ pairs. In our experiments with *p*-polarization aligned along the *a* axis, the 1.55 eV pump beam energy overlaps strongly with the broad 2 eV peak. We therefore expect the photons to be absorbed mainly due to intersite *d-d* transitions, with some smaller fraction of the absorption also coming from the tail of the charge transfer transitions.

Optical pump-probe studies have been directed at understanding the dynamics resulting from ultrafast excitation in $TbMnO_3$ as well as other related manganites. Pumping either the *p* → *d* charge transfer band [23] or the *d-d* transition peak [24] and probing via reflectivity reveals a sharp, time-resolution limited step observed only when the probe polarization was oriented along certain sample directions [24]. This fast step-like response is followed by a two-step relaxation process. The first is characterized by relaxation dynamics by time scales of several tens of picoseconds, very similar to what we recover here. The subsequent recovery to the initial state occurs much more slowly, with a time scale of several nanoseconds. Although the observed dynamics in the optical studies were interpreted in terms of time-dependent changes to magnetic properties, the current study has the clear advantage of directly monitoring the degree of magnetic order.

The next relevant step to understand the current measurements is to consider whether the two proposed excitation mechanisms directly modify the probed magnetic order. The *p* → *d* transition moves an electron from an oxygen site to a Mn site. The spin of the transferred electron is expected to be parallel to the total spin on the Mn site, as its antiparallel counterpart is energetically less favorable. The *p* → *d* charge transfer mechanism then results in an increase of the average Mn spin size as excited Mn sites change their spin from S=2 to S=2.5. This process therefore leads to a small *increase* in the average Mn magnetic scattering cross-section. Intersite Mn-Mn transitions, however, simply transfer spins from one Mn site to another, resulting in adjacent pairs of S=1.5 and S=2.5 sites. Here the average scattering cross-section along the measured (0*q*0) is unchanged. We therefore expect that, if anything, the immediate excitations of the sample would slightly increase the magnetic scattering structure factor. Since we observe no increase in the diffraction intensity, we conclude in particular that the *p* → *d* charge transfer mechanism presents only very small contribution to the absorption, consistent with our expectations from the optical dielectric function data. In this discussion we have assumed that there are no spectral changes in the x-ray resonance that could potentially also lead to a change of effective scattering cross-section. Since the experiment has been performed at the maximum of the $Mn^{3+}$ resonance, such changes could lead to a small reduction of the Mn scattering factor.

The intensity of resonant diffraction is directly proportional to the square of the average degree of long-range magnetic order $<S>^2$. Static measurements of the diffraction intensity as a function of temperature reveal that the diffraction intensity decreases roughly linearly with increasing temperature disappearing completely above $T_{N1}$ = ~42 K (see the inset to Figure 4(b) where the diffraction intensity is normalized to the 12 K value). A linear fit to the diffraction intensity for temperature up to $T_{N1}$ gives
$I/I_{(12\ K)} = 1.39(\pm 0.02) - 0.0311(\pm 0.001) \times T(K)$. Using this linear fit, we then convert the time-resolved diffraction intensity to an effective temperature of the spin system as shown in Figure 4(a). Note that we convert the data only over the range of relative intensities where the linear fit matched well the static data; we exclude later-time data above an effective temperature of 35 K for the highest fluences.

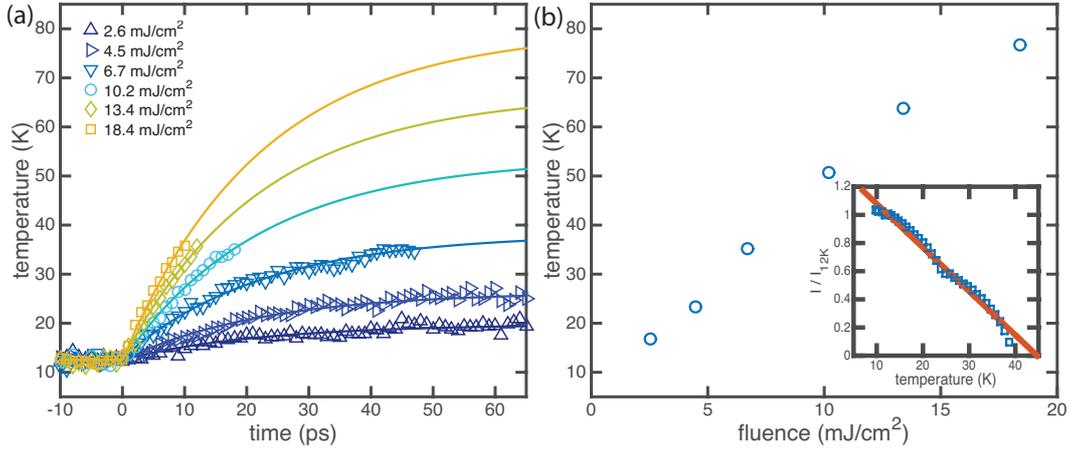

FIG 4. (Color Online) (a) Using static diffraction data (see the inset to (b)), we convert the time-resolved diffraction intensity to the temperature of the spin system. The lines are fits using Eq. (2) as described in the text. (b) Fits to Eq. (2) allow us to extract $T_f$, the temperature the spin system increases to, as a function of absorbed fluence. The inset shows the static diffraction intensity as a function of temperature, scaled to the intensity value at 12 K, taken at the same azimuthal angle (sample orientation). The line is a linear fit used to convert transient diffraction intensity to effective spin system temperature in (a).

We propose that the slow melting of magnetic order in this system is an indication of delayed transfer of thermal energy from the excited carriers to the spin system, where an effective spin temperature $T_{spin}$ characterizes the transferred energy. In Figure 3(b), we showed that a fit of the diffraction intensity to an exponential decay reveals a nearly constant decay time of $\tau$ = 22.3 ± 1.1 ps at low fluences. We posit that the effective spin temperature relaxes with a time dependence given by

$$T_{spin} = \begin{cases} T_0 & \text{for } t < 0 \\ (T_0 - T_f)\exp(-t/\tau) + T_f & \text{for } t > 0 \end{cases} \quad (2)$$

where $T_0$ is the initial temperature (12 K) and $T_f$ is the effective spin system temperature after energy transfer from the photoelectrons to the spin system via a relaxation mechanism. Using the fixed value of $\tau$ = 22.3 ps corresponding to the average extracted from the three lowest fluence traces, we fit the converted temperature data in Figure 4(a) using Eq. (2). The solid lines in Figure 4(a) are the fits where only the value of $T_f$ was allowed to vary. For all traces, there is good agreement using the fixed low fluence $\tau$. In Figure 4(b), we show the extracted $T_f$ as a function of fluence, which ranges from 19 K for the lowest fluence to 80 K for the highest. The nearly linear behavior (with perhaps a small departure at the highest fluence) shows that the spin system increases in temperature to a value directly proportional to the amount of energy deposited in the sample.

Although this effective temperature model reproduces the data well, the microscopic mechanisms for the melting process remain unclear. One possible relaxation pathway leading to the slow transfer of energy from the photoelectrons to the spin system could be tied to relaxation of the Jahn-Teller distortion. In other manganites, there is some theoretical and experimental evidence showing small polaron formation following disruption of the regular, pure $Mn^{3+}$ arrangements and relaxation of the Jahn-Teller mode due to optical excitation of $Mn^{2+}$ states [23,25,26]. This localization of the charge would hinder magnon-assisted hopping and therefore would require that the energy transfer to the spin system to be mediated by ultrafast changes in the lattice structure. Further time-resolved studies directly sensitive to lattice distortions, as have been done on other systems [27,28], would be required to confirm this possibility in $TbMnO_3$.

Another interesting observation is that the scattering wavevector of the spin-cycloid diffraction peak does not change significantly upon excitation, in contrast to changes observed in static experiments over corresponding temperature ranges. The static measurements show that the scattering wavevector changes from $q$ = 0.287, where it locks in at low temperatures, to $q$ = 0.30 at $T_N$ = 42 K at which the diffraction peak disappears [2]. This onset of the wavevector shift is also accompanied by a change in the long-range magnetic order at $T_{N2}$ = ~27 K, which reflects a phase transition from a cycloid to a sinusoidal spin structure. In the current measurements we see much smaller shifts in $q$: even for the highest pump fluence where we observe a nearly complete transient melting of the magnetic order, we measure a scattering wavevector shift from q = 0.287 to only 0.289 (see Figure 2(a)). When compared with static diffraction measurements, this shift corresponds to a temperature change of only a few degrees – very different from the spin system temperature rise to 80 degrees that we predict based on changes in diffraction intensity. This could indicate either that we melt the cycloidal phase and pass directly to the paramagnetic state, or that the material transforms first to the sinusoidal spin density wave state and then to the paramagnet, but without sufficient time to change the long-range magnetic ordering wavevector in the intermediate sinusoidal phase.

In the static case, it is understood that the magnitude of the spin order scattering wavevector is set by exchange coupling coefficients [29,30]. As the sample temperature increases, Monte Carlo simulations based on a classical Heisenberg model [30] predict the cycloidal to sinusoidal phase transition. Experimentally this is related to a shift in the scattering wavevector to larger values for increasing the temperature above this phase transition. The lack of a large change in scattering wavevector upon excitation with 1.55 eV light suggests that under these conditions, propagation of changes in the long-range ordering of the spin system within a single domain is very slow; indeed the magnon dispersion indicates an exceedingly low group velocity for all relevant wavevectors [5,6]. Thus even if the spin system is very quickly disrupted at the sample surface, it would necessarily take a long time to propagate this perturbation into the bulk. We might therefore expect that a fast increase in spin temperature would manifest itself first as a decrease in diffraction intensity and then only much later result in a change in the long-range order scattering wavevector.

## IV. Conclusions

Utilizing ultrafast near infrared excitation in tandem with soft x-ray resonant diffraction probing, we directly observe relatively slow melting of long-range magnetic order in $TbMnO_3$. We extract a magnetic order decay time constant of 22.3 ± 1.1 ps for low fluence pumping. Using this decay time, we additionally show that the spin system temperature increases linearly across our entire range of pump fluences; the pump laser excites the electronic system, which eventually transfers energy, heating the spin system. Whether this energy is mediated by the lattice due to the creation of a small polaron or simple lattice heating could be addressed in the future by probing other diffraction peaks sensitive to lattice and charge order [27]. Finally, the lack of large changes in the scattering wavevector indicates that despite the ultrafast pump, the extremely low magnon group velocity may prohibit fast changes in the spin-cycloid wavevector on these time scales.

This research was carried out on the SXR Instrument at the LCLS, a division of SLAC and an Office of Science user facility operated by Stanford University for the U.S. Department of Energy (DOE). The SXR Instrument is funded by a consortium including the LCLS, Stanford University through SIMES, Lawrence Berkeley National Laboratory, the University of Hamburg through the BMBF priority program FSP 301, and the Center for Free Electron Laser Science (CFEL). This research was supported by the National Center for Competence in Research (NCCR) Molecular Ultrafast Science and Technology and NCCR Materials with Novel Electronic Properties, funded by the Swiss National Science Foundation, and by the Swiss National Science Foundation (grant no. 200021_144115). Our ultrafast activities are supported by the ETH Femtosecond and Attosecond Science and Technology (ETH-FAST) initiative as part of the NCCR MUST program. J.A.J acknowledges support from the Marie Curie Actions IFP-MUST Cofund. The Advanced Light Source is supported by DOE under contract no. DE-AC02-05CH11231. Crystal growth work at IQM was supported by DOE, Office of Basic Energy Sciences, Division of Materials Sciences and Engineering under award DE-FG02-08ER46544. W.-S.L and R.G.M. are supported by DOE, Office of Basic Energy Sciences, Materials Sciences and Engineering Division, under contract DE-AC02-76SF00515. Y.-D.C. is supported by DOE, Office of Basic Energy Sciences, Materials Sciences and Engineering Division, under contract DE-AC02-05CH11231.  E. M. B acknowledges funding from the European Community's Seventh Framework Program (FP7/2007-2013) under grant agreement 290605 (COFUND: PSI-FELLOW).

*Correspondence to:  jjohnson@chem.byu.edu

†Current address: Department of Chemistry and Biochemistry, Brigham Young University, Provo, Utah, USA.